\documentstyle[12pt]{article}
\begin{document}

\begin{center}
{\Large \bf Universal Bound on Dynamical Relaxation Time
from Condition for Relaxing Quantity to be Classical}\\
\vskip .5cm
K. Ropotenko\\
\centerline{\it State Department of communications and
informatization} \centerline{\it Ministry of transport and
communications of Ukraine} \centerline{\it 22, Khreschatyk, 01001,
Kyiv, Ukraine}
\bigskip
\verb"ro@stc.gov.ua"

\end{center}
\bigskip\bigskip

\begin{abstract}
It is shown that the Hod's universal bound on the relaxation time of
a perturbed system \cite{hod} can be derived from a well-known
condition for a relaxing quantity to be classical in the fluctuation
theory.
\end{abstract}
\bigskip\bigskip

In a recent paper \cite{hod} Shahar Hod has derived the universal
bound on the relaxation time $\tau$ of a perturbed system from
information theory and thermodynamical considerations,
\begin{equation}
\label{tau}\tau\geq{\hbar/\pi T},
\end{equation}
where $T$ is the system's temperature.

In this note I want to point out that (\ref{tau}) follows from
quantum mechanics and thermodynamics directly without any reference
to information theory; it can be derived from a well-known condition
for a relaxing quantity to be classical in the fluctuation theory.

The proof is given in detail by Landau and Lifshitz \cite{land}. The
essential idea can be stated simply. The response of a system in
thermal equilibrium to an outside perturbation and the relaxation
timescale at which the perturbed system returns to an equilibrium
state can be considered in the framework of the fluctuation theory.
Landau and Lifshitz have found \cite{land} that if a fluctuating
quantity $x$ is to be classical, it must satisfy the condition:
\begin{equation}
\label{cond1} \tau\gg \hbar/T,~~~T\gg \hbar/\tau.
\end{equation}

This is just the Hod's universal bound. But here it is obtained in
more fundamental way and in that way, in our opinion, is more
appropriate. The condition (\ref{cond1}) ensures that quantum
effects are negligible in our thermodynamical considerations. But
when $\tau$ is too small ($x$ varies too rapidly) or when the
temperature is too low the fluctuations cannot be treated
thermodynamically, and the purely quantum fluctuations dominates.

In other words, the relaxing quantity $x$ behaves classically if its
components with frequency $\omega$ not satisfying the condition
$\hbar \omega \ll kT$ are  negligible. The contributions of
frequency $\omega$ to the dependence $x$ on time appear due to the
matrix elements of $x$ between the stationary states differing by
$\hbar\omega$. Hence, if the relaxing quantity $x$ is to be
classical, it must have negligible matrix elements between states,
the energy difference of which is not small as compared with $kT$.

That is why a typical laboratory system in the Hod's example has the
relaxation timescale which satisfies the more stronger condition
(\ref{cond1}) than (\ref{tau}). It is important to note that an
equilibrium state of a system is independent of how and under what
conditions this state was reached: due to the thermodynamical or
quantum fluctuations. Thus the Hod's bound holds in any case. This
is also clear from dimensional arguments.

But in case of a black hole the relaxation time is of the same order
of magnitude as the minimal relaxation time, $\tau_{min}=\hbar/T$.
In that case, as has been shown above, the quantum fluctuations are
large. Does it mean that a relaxing black hole parameter doesn't
behave classically? I think that this is not the case. First, as far
as is known the black hole perturbations are quite satisfactorily
governed by the classical Regge-Wheeler and Teukolsky equations.
Secondly, and what is more important, the point is that the time
$\tau$ need not be the same as the relaxation time for equilibrium
to be reached with respect to $x$, and may be less than this time if
$x$ approaches $<x>$ in an oscillatory manner \cite{land}. For
example, in case of the variation of pressure in a region of a body
with linear dimensions $R$, $\tau$ will be of the order of the
period of acoustic vibrations with wavelength $\lambda \sim R$, i.e.
$\tau \sim R/c$, where $c$ is the velocity of sound. The quasinormal
modes of a black hole have just the same oscillatory character.

\end{document}